\documentclass[aps,prd,preprint,nofootinbib,showpacs,preprintnumbers,amsmath,amssymb]{revtex4}

\usepackage{graphicx}% Include figure files
\usepackage{dcolumn}% Align table columns on decimal point
\usepackage{bm}% bold math

\def\lsim{\mathrel{\rlap{\lower4pt\hbox{\hskip1pt$\sim$}}
    \raise1pt\hbox{$<$}}}                % less than or approx. symbol
    
\begin{document}

\title{Gravitational-wave confusion background from cosmological
compact binaries: Implications for future terrestrial detectors}

\author{T.\ Regimbau}
\email{regimbau@oca.eu}
\affiliation{UMR ARTEMIS, CNRS, University of Nice Sophia-Antipolis,
Observatoire de la C\^ote d'Azur, BP 4229, 06304 Nice (France)}
\homepage{http://www.oca.eu/regimbau}

\author{Scott A.\ Hughes}
\affiliation{Department of Physics and MIT Kavli Institute, 77
Massachusetts Avenue, Cambridge, MA 02139}

\date{\today}

\begin{abstract}
Increasing the sensitivity of a gravitational-wave (GW) detector
improves our ability to measure the characteristics of detected
sources.  It also increases the number of weak signals that contribute
to the data.  Because GW detectors have nearly all-sky sensitivity,
they can be subject to a confusion limit: Many sources which cannot be
distinguished may be measured simultaneously, defining a stochastic
noise floor to the sensitivity.  For GW detectors operating at present
and for their planned upgrades, the projected event rate is
sufficiently low that we are far from the confusion-limited regime.
However, some detectors currently under discussion may have large
enough reach to binary inspiral that they enter the confusion-limited
regime.  In this paper, we examine the binary inspiral confusion limit
for terrestrial detectors.  We consider a broad range of inspiral
rates in the literature, several planned advanced gravitational-wave
detectors, and the highly advanced ``Einstein Telescope'' design.
Though most advanced detectors will not be impacted by this limit, the
Einstein Telescope with a very low frequency ``seismic wall'' may be
subject to confusion noise.  At a minimum, careful data analysis will
be require to separate signals which will appear confused.  This
result should be borne in mind when designing highly advanced future
instruments.
\end{abstract}

\maketitle

\section{\label{sec:intro}Introduction}

Compact binary coalescences, the gravitational-wave (GW) driven
inspiral and merger of binaries whose members are neutron stars and/or
black holes, are among the most promising sources of GWs for
ground-based detectors.  The late inspiral and merger waves from these
binaries are very strong, and their waveforms are predictable.  Their
detection would open a direct window into strong gravitational
dynamics, teaching us much about the evolution of massive stellar
systems, the nature of gravity in the strong-field regime, and perhaps
the nature of matter in neutron stars.  The well-modeled nature of
their waves may allow them to be used as standard candles (or standard
sirens) \cite{sch86,che93,fin96,hol05,aru07} to probe cosmology.  By
combining GW measurements, which encode distance to the source, with
optical or radio observations of the host galaxy (for example, if the
event is associated with a gamma-ray burst), which provides redshift,
one can calibrate the relation between luminosity distance and
redshift (the Hubble diagram) \cite{dal06}.

Such science goals have played a large role in driving plans for ever
more sensitive GW detectors.  Improved GW sensitivity impacts the
measurement of binary coalescence events in two ways.  First, any
measured event is characterized with greater precision by an
instrument with high sensitivity thanks to improved signal-to-noise
ratio.  Many of the analyses of GW science from compact binary
measurement assume measurement by ``advanced'' detectors (e.g.,
{\cite{fc93,cf94}}), with sensitivity to GWs roughly a factor of ten
or so greater than those in operation today.  Second, the ``detection
horizon'' to which events can be measured grows as sensitivity is
improved.  GW detector noise is normally quoted as a {\it strain
sensitivity}.  Doubling an instrument's sensitivity increases its
detection horizon by a factor of two, and hence the volume to which an
instrument is sensitive by a factor of eight.  The rate at which
events are measured can in principle be substantially increased by
relatively modest improvements in instrumental sensitivity.

Such improvement in sensitivity is on the one hand mandated by the
rarity of strong gravitational-wave events.  Extrapolation from the
observed binary pulsar systems in our galaxy and population synthesis
calculations each suggest that binary mergers occur roughly once every
hundred thousand years per Milky Way equivalent galaxy; a more precise
discussion of the merger rate is given in Sec.\ {\ref{sec:rate}}
below.  Combining this rate with the density of galaxies in our
universe drives us to the need for a detection horizon of a few
hundred Mpc in order to measure multiple NS-NS events per year.  On
the other hand, there can be too much of a good thing: If the
detection horizon is sufficiently far out, sources may be detected so
often that they overlap, making a transition to a confusion-limited
background.  The transition from discrete sources to confused
background has been discussed extensively, especially in the context
of sources for the space-based detector LISA {\cite{phinney01,fp03}}.

In this paper, we investigate the threshold redshift at which merging
binary sources for ground-based detectors begin to overlap and result
in a confusion background.  Beyond this redshift, it will become
difficult to resolve sources individually.  As has been found in the
LISA Mock Data Challenges, separation may be possible by globally
fitting for all sources simultaneously (see, e.g., Ref.\
{\cite{babak08}}); similar analyses have been done for the proposed
highly-advanced Big Bang Observatory {\cite{cut06}}.  Careful analysis
will be needed to see how well such techniques can be applied to
ground-based detector data.  At any rate, it is likely that improving
sensitivity to probe beyond this point will not be worthwhile.  We
first discuss estimates of the cosmic coalescence rates of neutron
star-neutron star (NS-NS) and neutron star-black hole (NS-BH) systems
in Sec.\ {\ref{sec:rate}}.  In Sec.\ {\ref{sec:DC}}, we then discuss
the different detection regimes which pertain to binary coalescence
measurement.  Key to this discussion is understanding how long
different sources are in band, which depends most strongly on the
low-frequency sensitivity of the instrument being used to measure the
waves.  We consider the initial and planned advanced LIGO and Virgo
detectors, and three possible configurations of the {\it Einstein
Telescope} concept.  Our conclusions are given in Sec.\
{\ref{sec:conc}}.  Chief among them is that confusion issues are
likely to become important for ground-based GW detectors which are
sensitive to these events to a redshift $z \sim 1$, especially if such
instruments have a ``seismic wall'' at or below roughly 5 Hz.  This
may have important ramifications for designing highly sensitive future
detectors such as the Einstein Telescope.  At a minimum, it will be
necessary to carefully design the data analysis to disentangle these
potentially confused sources.

\section{\label{sec:rate}Coalescence rates}

To motivate the choices for our calculations, we begin by reviewing
our current understanding of the rate of NS-NS and NS-BH coalescences
in the universe.  The final merger of a compact binary occurs after
two massive stars in a binary system have collapsed to form neutron
stars or black holes\footnote{We neglect the possible production of
compact binaries through interactions in dense star systems.} and have
inspiralled through the emission of gravitational waves.  We assume
that the coalescence rate tracks the star formation rate, albeit with
some delay $t_d$ from formation of the binary to final merger.
Accordingly, we put
\begin{equation}
\dot\rho_c^{\rm o}(z) = \dot\rho_c^{\rm o}(0) \times
\frac{\dot\rho_{*,c}(z)}{\dot\rho_{*,c}(0)}\;.
\label{eq:dotrho_c}
\end{equation}
In Eq.\ (\ref{eq:dotrho_c}), $\dot\rho_c^{\rm o}(z)$ represents the
rate at which binary systems are observed to merge at redshift $z$,
and $\dot\rho_c^{\rm o}(0)$ is this rate in our local universe.  This
normalization reproduces the local rate for $z = 0$.  We review
estimates of $\dot\rho_c^{\rm o}(0)$ in more detail later in this
section.  The quantity $\dot\rho_{*,c}$ relates the past star
formation rate (SFR) to the rate of binary merger.  It is defined as
\begin{equation}
\dot\rho_{*,c}(z) = \int \frac {\dot\rho_*(z_f)} {(1+z_f) } P(t_d)
dt_d\;.
\label{eq:dotrho_star_c}
\end{equation}
In Eq.\ (\ref{eq:dotrho_star_c}), $\dot\rho_*$ is the SFR, measured in
M$_\odot$ Mpc$^{-3}$ yr$^{-1}$.  The redshift $z$ describes when our
compact binary merges, and $z_f$ is the redshift at which its
progenitor binary formed.  These redshifts are connected by the time
delay $t_d$, which is the sum of the time from initial binary
formation to evolution into compact binary, plus the merging time
$\tau_m$ by emission of gravitational waves.  It is also the
difference in lookback times between $z_f$ and $z$:
\begin{equation}
t_d = t_{\rm LB}(z_f) - t_{\rm LB}(z) = \frac{1}{H_0}
\int_z^{z_f} \frac{dz'}{(1 + z')E(\Omega, z')}\;,
\end{equation}
where
\begin{equation}
E(\Omega,z)=\sqrt{\Omega_{\Lambda}+\Omega_{m}(1+z)^3}\;.
\label{eq-E-z}
\end{equation}
We use the 737 cosmology \cite{rao06}, with $\Omega_m=0.3$,
$\Omega_{\Lambda}=0.7$ and Hubble parameter $H_0=70$ km s$^{-1}$
Mpc$^{-1}$ (or equivalently $h_0=0.7$).  This corresponds to the
concordance model derived from observations of distant type Ia
supernovae \cite{per99} and the power spectrum of the cosmic microwave
background fluctuations \cite{spe03}.  Finally, in Eq.\
(\ref{eq:dotrho_star_c}) $P(t_d)$ is the probability distribution for
the delay $t_d$.  The factor $1/(1+z_f)$ accounts for time dilatation
due to the cosmic expansion.

We find the merger rate per unit redshift by multiplying
$\dot\rho_c^{\rm o}(z)$ with the element of comoving volume:
\begin{equation}
\frac{dR_c^{\rm o}}{dz}(z)=\dot\rho_c^{\rm o}(z) \frac{dV}{dz}(z)\;,
\label{eq-dRdz}
\end{equation}
where
\begin{equation}
\frac{dV}{dz}(z)=4 \pi \frac{c}{H_0} \frac{r(z)^2}{E(\Omega,z)}\;,
\label{eq-dVdz}
\end{equation}
\begin{equation}
r(z)= \frac{c}{H_0}\int_0^z \frac{dz'}{E(\Omega, z')}\;.
\label{eq-r_z}
\end{equation}
Note that Eq.\ (\ref{eq-r_z}) assumes spatial flatness, $\Omega_m +
\Omega_\Lambda = 1$.

\begin{figure}
\centering
\includegraphics[angle=0,width=0.8\columnwidth]{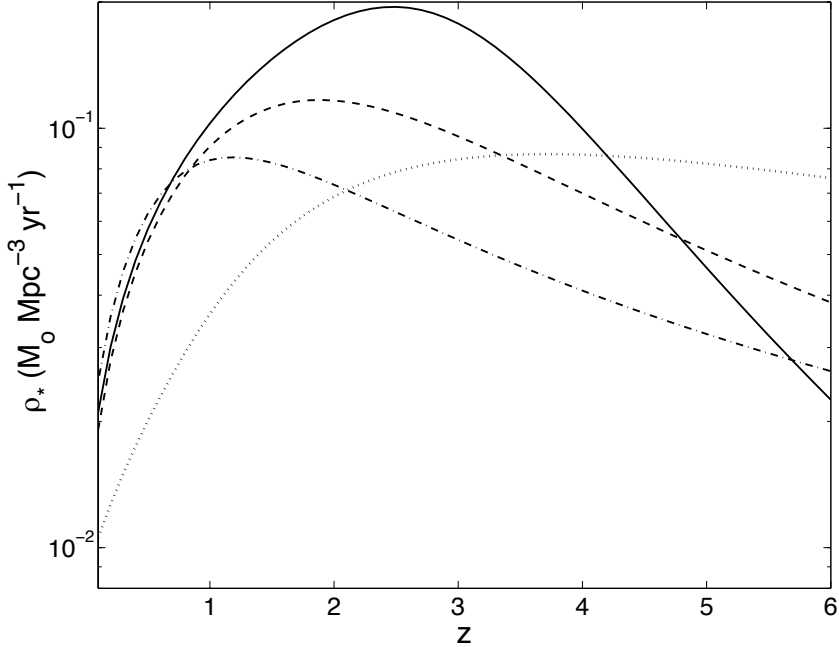}
\caption{Cosmic star formation rates (in M$_\odot$ Mpc$^{-3}$
yr$^{-1}$) used in this paper: Hopkins and Beacom 2006 \cite{hop06}
(continuous line), Fardal et al. 2007 \cite{far07} (dashed line),
Wilkins et al. 2008 (dot-dashed line), \cite{wil08}, and the fossil
model of Nagamine et al. 2006 (dot line).  As discussed in the text,
these rates are largely the same up to $z \sim 1$, but show important
differences at higher redshift.
\label{fig-allsfr}}
\end{figure}

For $\dot\rho_*$, we use the recent SFR of \cite{hop06}, which is
derived from new measurements of the galaxy luminosity function in the
UV (SDSS, GALEX, COMBO17) and FIR wavelengths (Spitzer Space
Telescope), and is normalized by the Super Kamiokande limit on the
electron antineutrino flux from past core-collapse supernovas.  This
model is expected to be quite accurate up to $z \sim 2$, with very
tight constraints at redshifts $z<1$ (to within $30-50 \%$).  To
account for uncertainty in reconstructing the SFR, we also considered
the model described in Fardal et al.\ (Ref.\ \cite{far07}).  That work
uses a different set of measurements and a different dust extinction
correction.  The SFR found in \cite{far07} is the same as that of
\cite{hop06} up to $z \sim 1$, but decreases slightly at higher
redshifts.  We also consider the model described by Wilkins et al.\ in
Ref.\ \cite{wil08}, which is derived from measurements of the stellar
mass density.  The SFR is equivalent to that in \cite{hop06,far07} for
$z \lesssim 0.7$, but again is lower at higher redshifts.  Finally, we
consider the SFR of Ref.\ \cite{nag06}, which is derived from the
fossil record of star formation in nearby galaxies.  It is probably
underestimated at small redshifts, and is constant at high redshifts
due to the contribution of elliptical galaxies.  Note that at present
there is a discrepancy between the ``instantaneous'' SFR, measured
from the emission of young stars in star forming regions, and the SFR
as determined from extragalactic background light.  This could have an
important impact on the contribution to the confusion background for
sources from $z > 2$.

Population synthesis \cite{pir92,tut94,lip95,dfp06,and04,sha08,bel06}
suggests that the delay time $t_d$ is well described by a probability
distribution of the form
\begin{equation}
P_d(t_d) \propto \frac{1}{t_d} \;\ \mathrm{with} \;\ t_d > \tau_0
\end{equation}
for some minimal delay time $\tau_0$.  This broad model accounts for
the wide range of merger times observed in binary pulsars; it is also
consistent with short gamma ray burst observations in both late and
early type galaxies \cite{ber06}.  Following \cite{bel01,bel06}, who
identify a channel which produces tight NS-NS binaries with merger
time in the range $\tau_m \sim 0.001-0.1$ Myr, we assume a minimal
delay time for NS-NS of $\tau_0 \sim 20$ Myr.  This corresponds
roughly to the time it takes for massive binaries to evolve into two
neutron stars.  For NS-BHs, we take a minimal delay of $\tau_0 \sim
100$ Myr \cite{bul04}, corresponding to the wider orbits and longer
merger times predicted by classical channels.

\begin{table}
\caption{\label{table-rates} Current estimates of the Galactic merger
rates of NS-NSs and NS-BHs in units of Myr$^{-1}$, adapted from Table
4 of \cite{pos06}.  The rates from Ref.\ {\cite{kal04}} are derived
from statistical studies; the numbers in parenthesis give the 95\%
confidence limits around the maximum likelihood value.  All other
values are estimated using population synthesis. The high rate
obtained in Ref.\ {\cite{tut93}} is due to the assumption that neutron
stars or black holes are born with no kick velocity, leading to an
overestimate of the number of systems that survive both supernovae.
The low rate obtained by Ref.\ {\cite{vos03}} is due to their
treatment of common envelope binding physics.}
\begin{ruledtabular}
\begin{tabular}{lcc}
\bf{Statistics} & NS-NS &\\
Kalogera et al.\ (2004) {\cite{kal04}} & 83 (17 -- 292) &\\
\hline
\bf{Population synthesis} & NS-NS &  NS-BH \\
Tutunov and Yungelson (1993) {\cite{tut93}} & 300 & 20 \\
Lipunov et al.\ (1997) {\cite{lip97}} & 30 & 2 \\
Portegies Zwart and Yungelson (1998) {\cite{por98}} & 20 & 2 \\
Nelemans et al.\ (2001) {\cite{nel01}} & 20 & 4 \\
Voss and Tauris (2003) {\cite{vos03}} & 2 & 0.6 \\
de Freitas Pacheco et al.\ (2006) {\cite{dfp06}} & 17  \\
Belczinsky et al.\ (2007) {\cite{bel07}} & 10-15 & 0.1 \\
O'Shaughnessy et al.\ (2008) {\cite{sha08}} & 30 & 3 \\
\end{tabular}
\end{ruledtabular}
\end{table}

The local merger rate per unit volume, $\dot\rho_c^{\rm o}(0)$, is
usually extrapolated by multiplying the rate in the Milky Way
($r_{\mathrm{mw}}$) with the density of Milky-Way equivalent galaxies.
That density, in turn, is found by measurements of the blue stellar
luminosity to be roughly $n_{\mathrm{mw}} \sim (1-2) \times 10^{-2}$
Mpc$^{-3}$ \cite{phi91,kal01,kop08}.  Current estimates of the NS-NS
galactic coalescence rate which extrapolate from observed galactic
NS-NS find a rate in the range $17-292$ Myr$^{-1}$ ($95\%$ confidence
interval), with a most likely rate of 83 Myr$^{-1}$ \cite{kal04}.
Population synthesis predicts a NS-NS merger rate in the range $1-300$
Myr$^{-1}$ (most likely near $10-30$ Myr$^{-1}$; see Table
\ref{table-rates} for a summary).  Rate estimates for NS-BH systems
are one or two orders of magnitudes smaller, ranging between $0.1-30$
Myr$^{-1}$ (Tab.\ \ref{table-rates}).

It is worth noting the stellar mass fraction which produces massive
binaries is expected to be larger in early-type galaxies than in
spiral galaxies thanks to their flatter initial mass function.
Despite their absence of recent star formation, early-type galaxies
may make a larger contribution to the merger rate than spiral galaxies
due to binaries with very long coalescence times which were born in a
galaxy's first $1-2$ Gyr.  Assuming that ellipticals represent a
fraction of about $40 \%$ of the galaxies, \cite{dfp06} find a small
correction factor $\sim 2$.

Taking these uncertainties into account, we consider local rates in
the range $\dot\rho_c^{\rm o}(z = 0) =(0.01-10)$ Myr$^{-1}$ Mpc$^{-3}$
for NS-NS and $\dot\rho_c^{\rm o}(z = 0) = (0.001-1)$ Myr$^{-1}$
Mpc$^{-3}$ for NS-BH.  Our two reference models for NS-NS are
\begin{itemize}

\item $\dot\rho_c^{\rm o}(z = 0) \sim 1$ Myr$^{-1}$ Mpc$^{-3}$,
corresponding to the most probable galactic rate of $r_{\mathrm{mw}} =
83\,{\rm Myr}^{-1}$, and

\item $\dot\rho_c^{\rm o}(z = 0) \sim 0.4 $ Myr$^{-1}$ Mpc$^{-3}$,
corresponding to the prediction of the latest population synthesis of
\cite{sha08} ($r_{\mathrm{mw}}=30$ Myr$^{-1}$ or to
$r_{\mathrm{mw}}=15$ Myr$^{-1}$ \cite{dfp06,bel07}, with a correction
factor of 2 due to ellipticals).

\end{itemize}
Our reference model for NS-BH is $\dot\rho_c^{\rm o}(z = 0) \sim 0.04
$ Myr$^{-1}$ Mpc$^{-3}$, corresponding to the most recent predictions
of \cite{sha08b}.  For all three models, we assume $n_{\mathrm{mw}}=
1.2 \times 10^{-2}$ Mpc$^{-3}$ \cite{kop08}.

\begin{figure}
\centering
\includegraphics[angle=0,width=0.8\columnwidth]{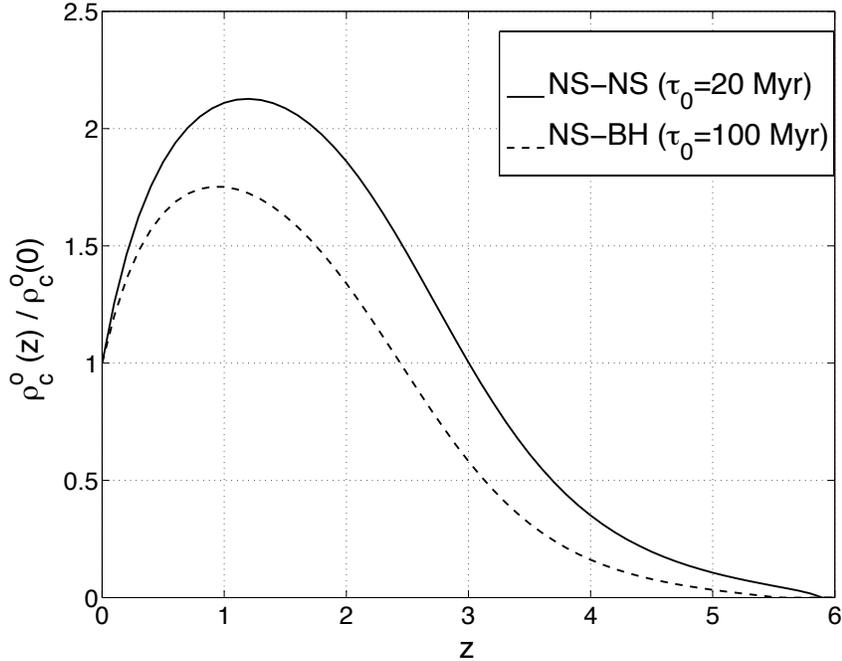}
\caption{Cosmic coalescence rate, normalized to the local value
$\dot\rho_c^{\rm o}(z = 0)$, for models with a distribution of the
delay of the form $P(t_d) \propto 1/t_d$.  The continuous line is the
distribution for a minimal delay $\tau_0 = 20$ Myr, which is assumed
to be representative for NS-NS binaries.  The dashed is for a
delay of 100 Myr, taken to be representative of NS-BHs.  In both
cases, we have assumed the cosmic star formation rate of Hopkins and
Beacom \cite{hop06}.  
\label{fig-sfr}}
\end{figure}

\section{\label{sec:DC}Detection regimes}

Turn now to the measurement of signals from NS-NS and NS-BH binaries.
The contribution of these binaries to the instrumental data falls into
three statistically very different regimes, depending mostly on the
typical interval between events (see \cite{dra03,cow06} and references
within):

\begin{enumerate}

\item {\it Shot noise}: This case describes when the number of sources
is small enough that the interval between events is long compared to
an individual event's duration.  Measured waves are separated by long
stretches of silence and can be resolved individually.  This case
pertains to instruments that are only sensitive to events at low
redshift.

\item {\it Popcorn noise}: As the reach of instruments increases, the
time interval between events may come closer to the duration of single
bursts.  Events may sometimes overlap, making it difficult to
distinguish between them.

\item {\it Continuous}: For instruments with very large reach and
excellent low frequency sensitivity, the interval between events can
be small compared to the duration of an event.  The signals overlap to
create a confusion noise of unresolved sources.

\end{enumerate}

The average number of measured events at a given moment is given by
the duty cycle $\Delta$.  It is defined as the ratio, in the observer
frame, of the typical duration of a single event $\tau$, to the
average time interval between successive events
\begin{equation}
\Delta(z)=\int^z_0 \tau(z') \frac{dR_c^{\rm o}}{dz'}(z') dz'\;.
\label{eq-DC}
\end{equation}
Here, $dR_c^{\rm o}/dz' \propto \dot\rho_c^{\rm o}$ is the coalescence rate
per unit redshift, given by Eq.\ (\ref{eq-dRdz}); $\tau(z')$ is
the observed duration of a GW signal generated at redshift $z'$.  This
duration is given at leading order by
\begin{equation}
\tau(z') = \frac{5 c^5}{256 \pi^{8/3}G^{5/3}} [(1 + z')m_c]^{-5/3}
f_L^{-8/3}\;,
\label{eq-tau}
\end{equation}
where $m_c = (m_1 m_2)^{3/5}/(m_1 + m^2)^{1/5}$ is the binary's {\it
chirp mass}, and where $f_L$ is the lower frequency bound of the
detector, assumed to be much smaller than the frequency at the time of
the merger.

\begin{figure}
\centering
\includegraphics[angle=0,width=0.8\columnwidth]{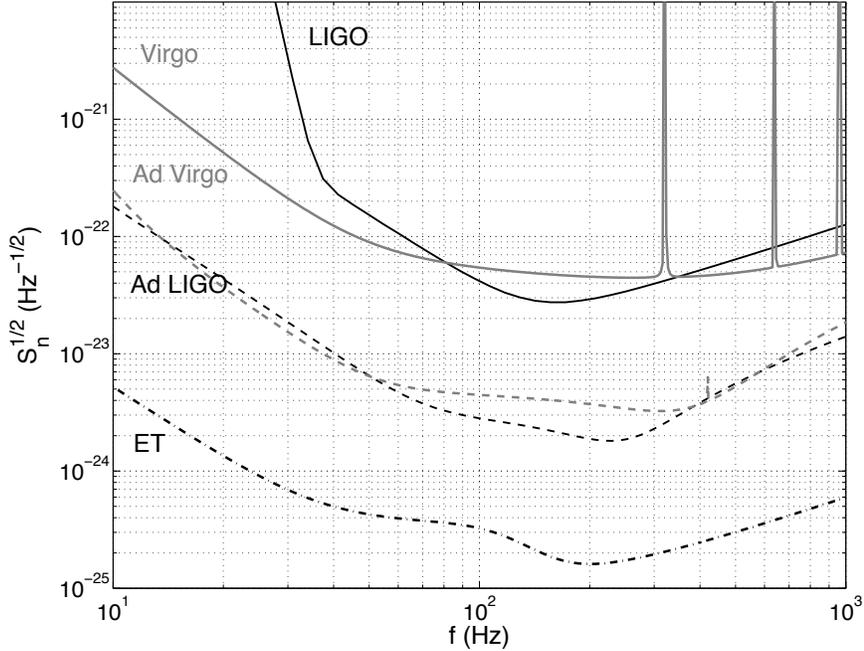}\\
\caption{Strain sensitivities, in ${\rm Hz}^{-1/2}$, for the
instruments that we consider as functions of frequency: LIGO 4 km as
specified in the Science Requirement Document \cite{laz96}; design
initial Virgo \cite{pun04}; planned Advanced LIGO and Virgo
\cite{fri03,adL08,adV08} (optimized for NS-NS detection); and a
possible Einstein Telescope design (L-shaped, 10 km) [ET science team,
private communication].
\label{fig-sens}}
\end{figure}

The signal duration depends very strongly on both chirp mass and lower
frequency bound; in particular, signals measured by instruments with
very low $f_L$ may have very long durations.  We will take typical NS
masses to be 1.4 M$_\odot$, and typical BH mass to be 9.5 M$_{\odot}$,
yielding chirp masses $m_c = 1.2$ M$_{\odot}$ for NS-NS and $m_c =
2.9$ M$_{\odot}$ for NS-BH.  These values agree well with the most
probable values derived by \cite{bel07} with the StarTrack population
synthesis code, for their reference model.  The lower frequency bound
is determined by the properties of the detector used for the
measurement.  Sensitivity curves describing the instruments we include
in our analysis are shown in Fig.\ {\ref{fig-sens}}. 
We consider:

\begin{itemize}

\item {\it Initial LIGO}: The present LIGO interferometers
\cite{laz96} have a low-frequency seismic ``wall'' at roughly 40 Hz.
NS-NS binaries are in band for roughly 25 seconds; NS-BH binaries are
in band for roughly 5.8 seconds.  The isotropic detection horizon (the
angle-averaged distance to which a binary can be measured, defined
carefully below) is at a luminosity distance of 15 Mpc ($z = 0.0035$)
for NS-NS, 30 Mpc for NS-BH ($z = 0.007$) for the LIGO Hanford 4
kilometer detector.  When both 4 kilometer detectors (at Hanford,
Washington and Livingston, Louisiana) and the single 2 kilometer
detector (at Hanford) are combined (assuming independence of all
detectors and stationary noise), these numbers increase by a factor of
about $\sqrt{1^2 + 1^2 + (1/2)^2} = 1.5$ to 22 Mpc ($z = 0.005$) for
NS-NS, 44 Mpc for NS-BH ($z = 0.01$).

\item {\it Virgo}: The Virgo design sensitivity \cite{pun04} is
comparable to the LIGO sensitivity, but with advanced seismic
isolation provided by the so-called ``superattenuator'' (see, e.g.,
Ref.\ \cite{fab87} for discussion).  A low-frequency wall at 10 Hz
means that NS-NS signals will be in band for 16.7 minutes; NS-BH
signals last for 3.9 minutes.  The Virgo sensitivity is slightly
poorer than the LIGO sensitivity between $100-300$ Hz, so its
isotropic detection horizon is 11 Mpc ($z = 0.0025$) for NS-NS, 23 Mpc
for NS-BH ($z = 0.0055$).

\item {\it Advanced LIGO/VIRGO}: The advanced LIGO design
\cite{fri03,adL08} should move its wall down to a frequency of roughly
10 Hz.  NS-NS signals will be in band for about 16.7 minutes and have
a single detector detection horizon of 200 Mpc ($z = 0.045$); NS-BH
signals will be in band for 3.9 minutes, and have a detection horizon
of 420 Mpc ($z = 0.09$).  The current Advanced LIGO design plan will
have three 4 kilometer instruments; combining their data brings the
isotropic horizon to 355 Mpc ($z = 0.08$) for NS-NS and 765 Mpc ($z =
0.16$) for NS-BH.  The plans for advanced Virgo sensitivity
\cite{adV08} are comparable to those for LIGO. The expected isotropic
detection horizon for advanced Virgo is at 150 Mpc ($z = 0.035$) for
NS-NS, 310 Mpc for NS-BH ($z = 0.07$).

\item {\it The Einstein Telescope}: Several possible Einstein
Telescope designs are presently under discussion.  The plan is to move
the seismic wall down to a frequency below 5 Hz.  A planned
low-frequency wall at 5 Hz would mean that NS-NS signals will be in
band for 1.8 hours; NS-BH signals will be for 24.6 minutes.  Another
possibility will be to lower it to 3 Hz (so that NS-NS signals are in
band for 6.9 hours, NS-BH for 1.6 hours), or even to 1 Hz (NS-NS
lasting for 5.4 days, NS-BH for 1.2 days).  The Einstein Telescope
detection horizon for NS-NS signals is at $z \simeq 1$; for NS-BH, it
is at $z \simeq 2$.
\end{itemize}

The isotropic detection horizon for these instruments is the distance
at which the angle-averaged signal-to-noise ratio $\rho$, defined by
\begin{equation}
\rho^2 = 4\left\langle\int \frac{|{\tilde
h}(f)|^2}{h_n(f)^2}df\right\rangle\;,
\label{eq:snrsquared}
\end{equation}
equals the threshold value $\rho_{\rm th} = 8$.  In Eq.\
(\ref{eq:snrsquared}), $\tilde h(f)$ is the Fourier transform of the
inspiral waveform, $h_n(f)$ is the strain noise (in units of ${\rm
Hz}^{-1/2}$, as shown in Fig.\ {\ref{fig-sens}}), and $\langle
\rangle$ denotes an appropriate angle averaging; see, for example,
Ref.\ {\cite{dal06}}, Sec.\ II for discussion.  We are careful to
specify the ``isotropic'' detection horizon in our discussion (in
reference to this angle averaging) to contrast with the term
``detection horizon.''  This name is typically used to denote the
distance to which an {\it optimally} oriented and position source
could be detected.  Whereas fortunate orientation and position may
make a source visible beyond the {\it isotropic} detection horizon,
{\it no} source is visible beyond the detection horizon.  The
threshold $\rho_{\rm th} = 8$ is chosen to keep the false alarm rate
(for Gaussian noise statistics) acceptably low.

The redshift $z_*$ at which sources start to overlap to produce a
confusion background is the transition between the shot noise and the
popcorn noise regimes.  This is defined by the condition
\begin{equation}
\Delta(z_*)=1\;.
\label{eq-thresh}
\end{equation}
In this regime, where the number of sources that overlap is small, it
may still be possible to separate the sources individually with
adequate data analysis techniques.  We therefore consider in addition
a more conservative threshold $z_{**}$, corresponding to the
transition between the popcorn regime and the Gaussian stochastic
background.  It is defined by $\Delta(z_{**}) = 10$.

\begin{figure}
\centering
\includegraphics[angle=0,width=0.48\columnwidth]{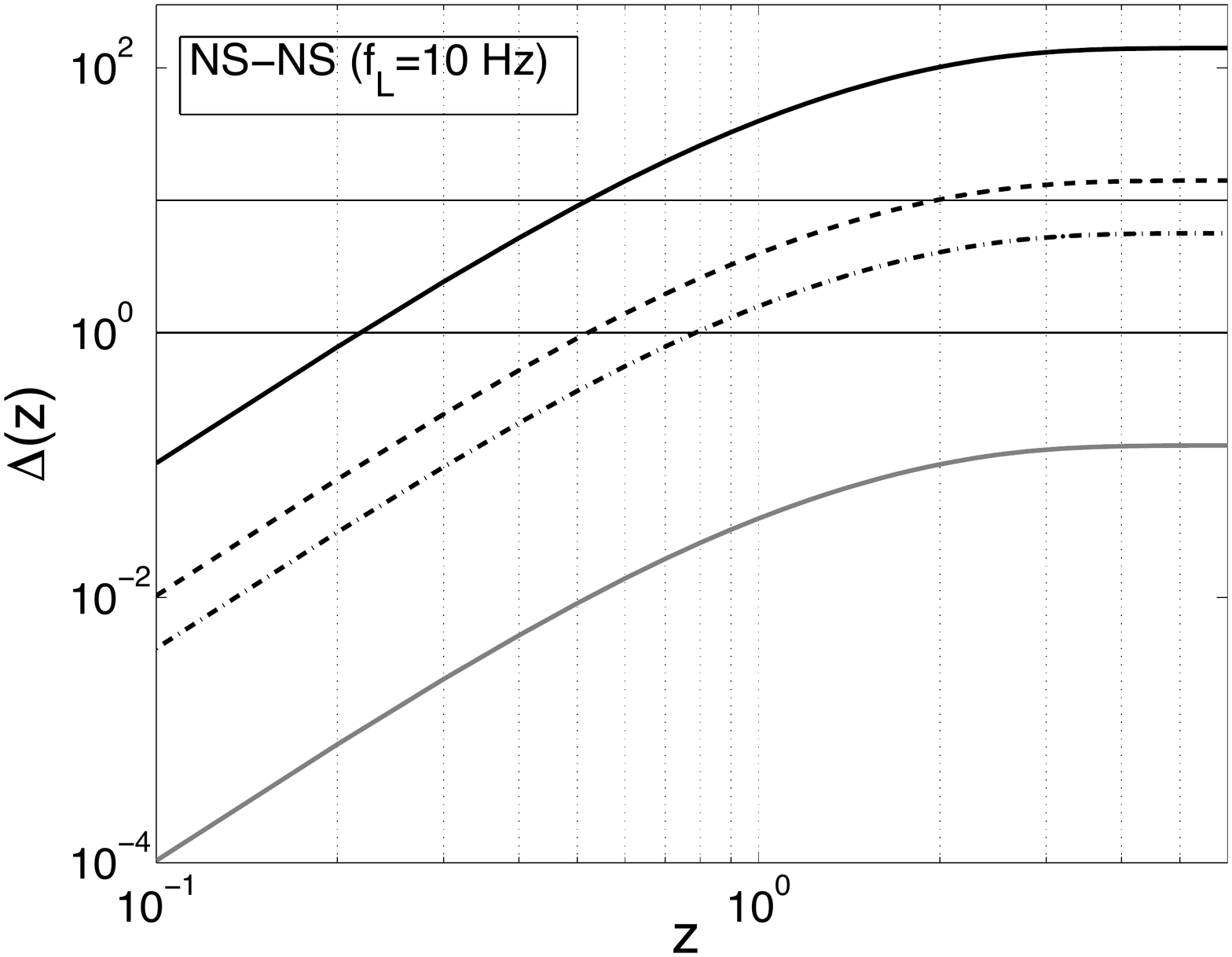}
\includegraphics[angle=0,width=0.48\columnwidth]{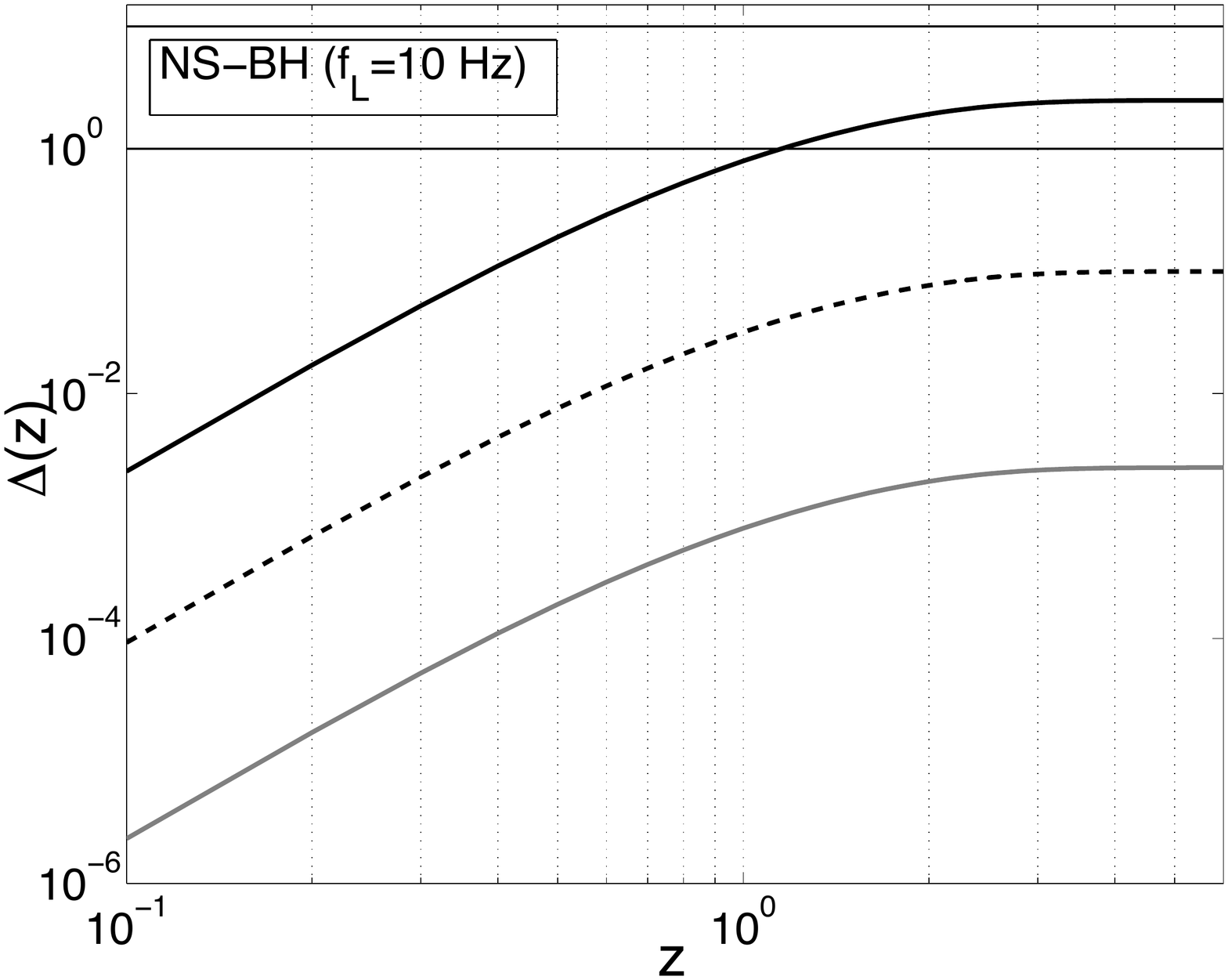}
\caption{Duty cycle as a function of redshift for NS-NSs (top) and
NS-BHs (bottom), for a lower frequency bound of 10 Hz and for our
reference SFR \cite{hop06} . The continuous black and grey lines
correspond to the most optimistic and pessimistic estimates of the
local coalescence rate, respectively; the dashed and dot-dash lines
correspond to our reference models (cf.\ Sec.\ {\ref{sec:rate}}).  The
horizontal line at $\Delta(z) = 1$ indicates the transition between
resolved sources and a ``popcorn'' background; the line at $\Delta(z) = 10$
is our more conservative transition to a confused background.
\label{fig-DC10}}
\end{figure}  
\begin{figure}
\centering
\includegraphics[angle=0,width=0.48\columnwidth]{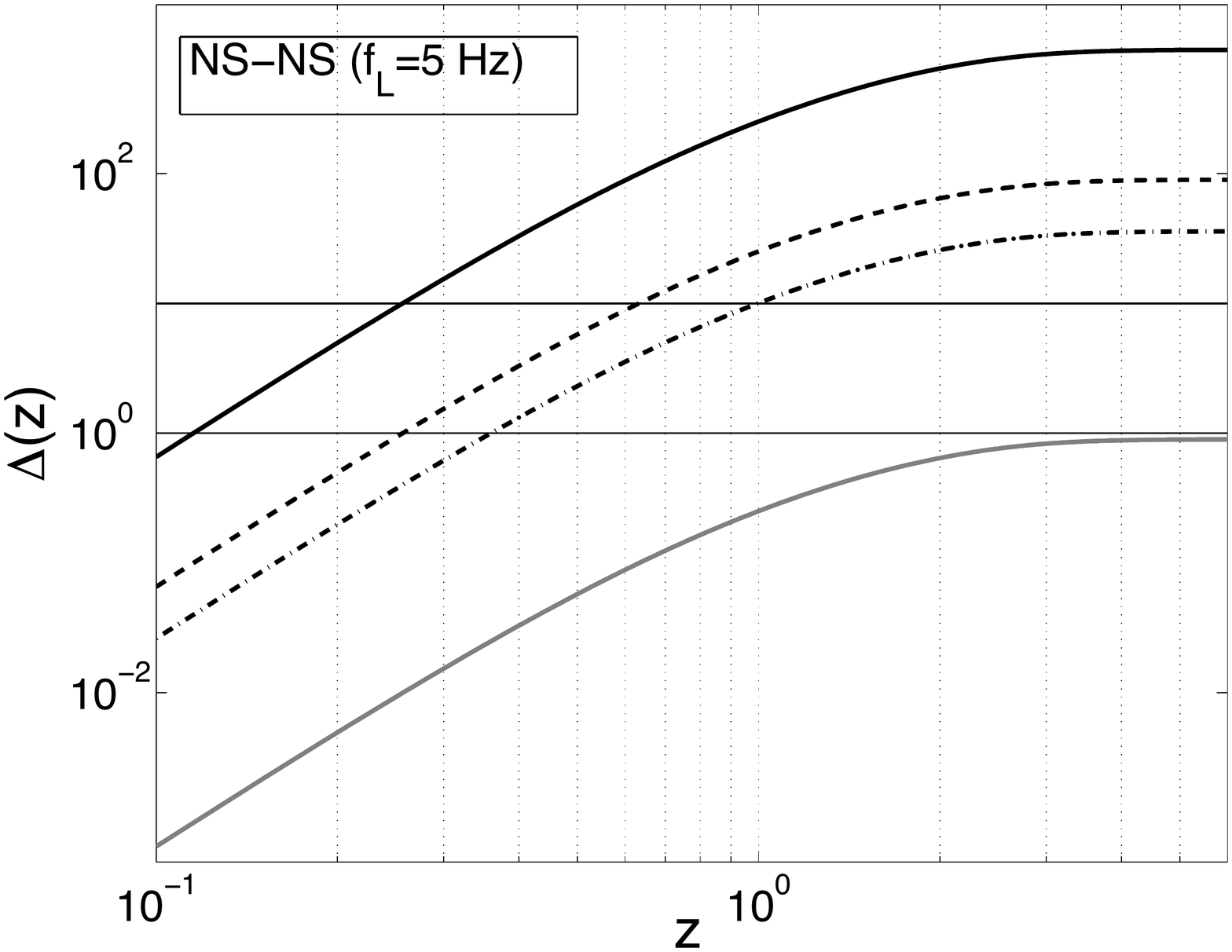}
\includegraphics[angle=0,width=0.48\columnwidth]{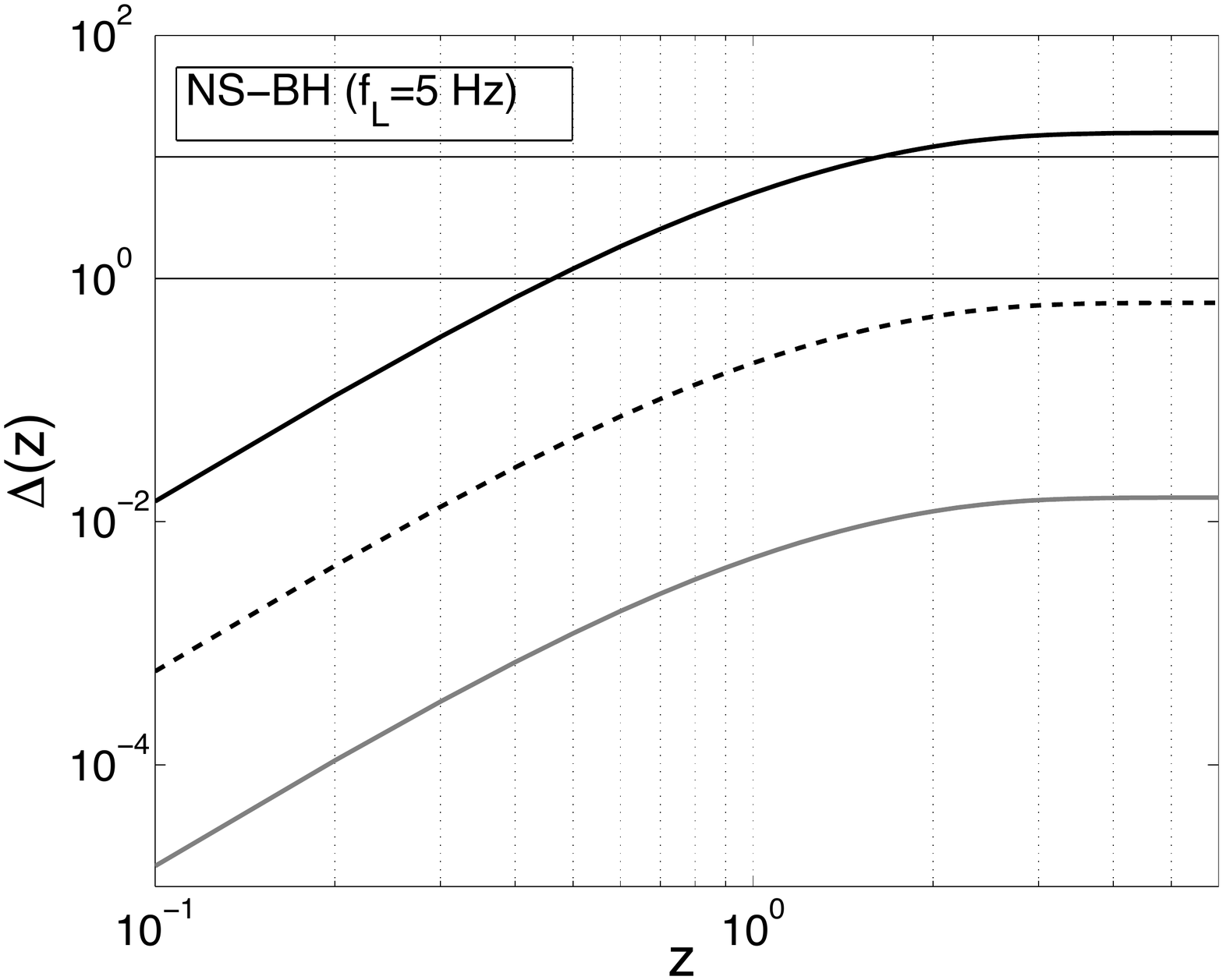}
\caption{Same as Fig.~\ref{fig-DC10} for a lower frequency bound of 5
Hz.
\label{fig-DC5}}
\end{figure}  
\begin{figure}
\centering
\includegraphics[angle=0,width=0.48\columnwidth]{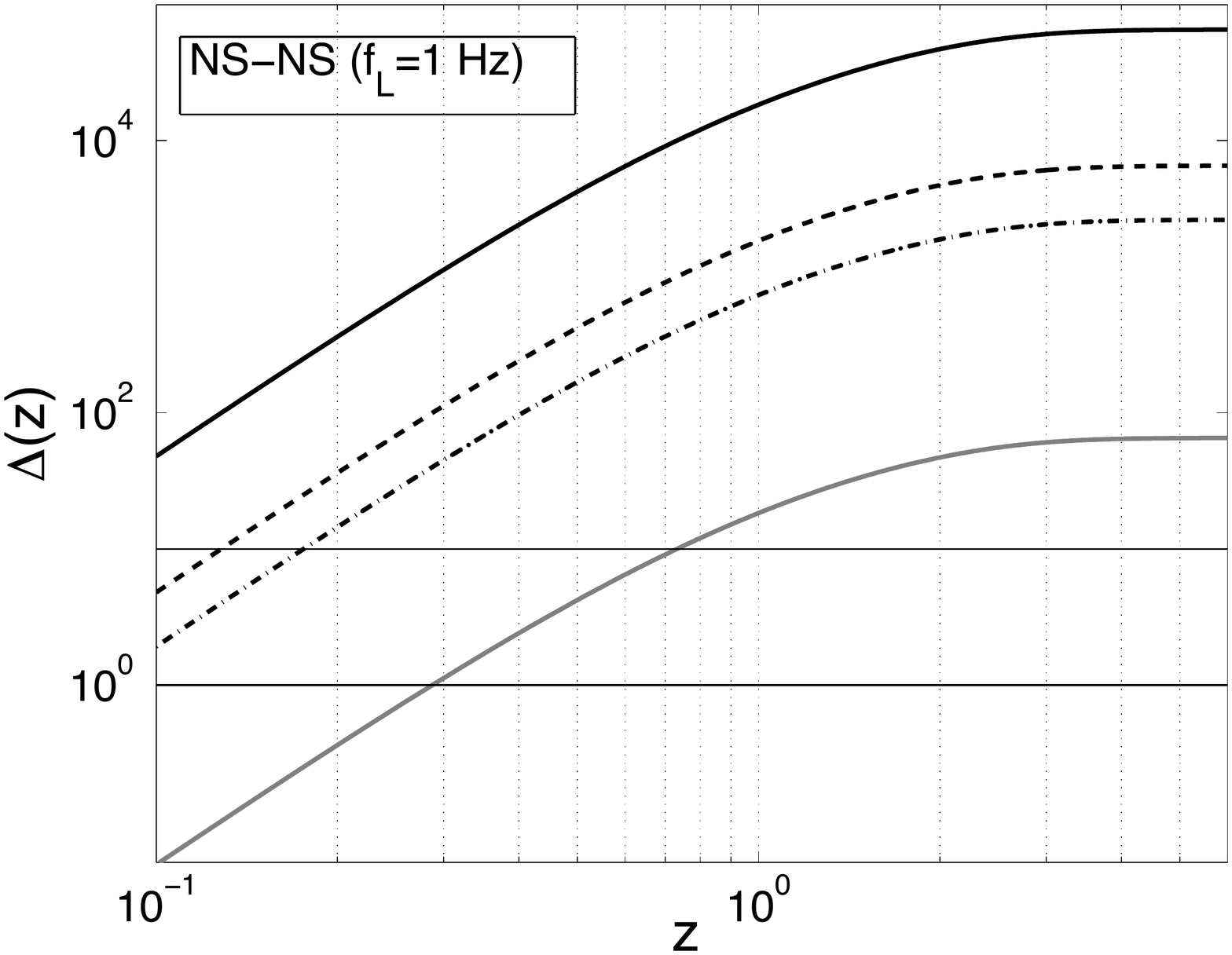}
\includegraphics[angle=0,width=0.48\columnwidth]{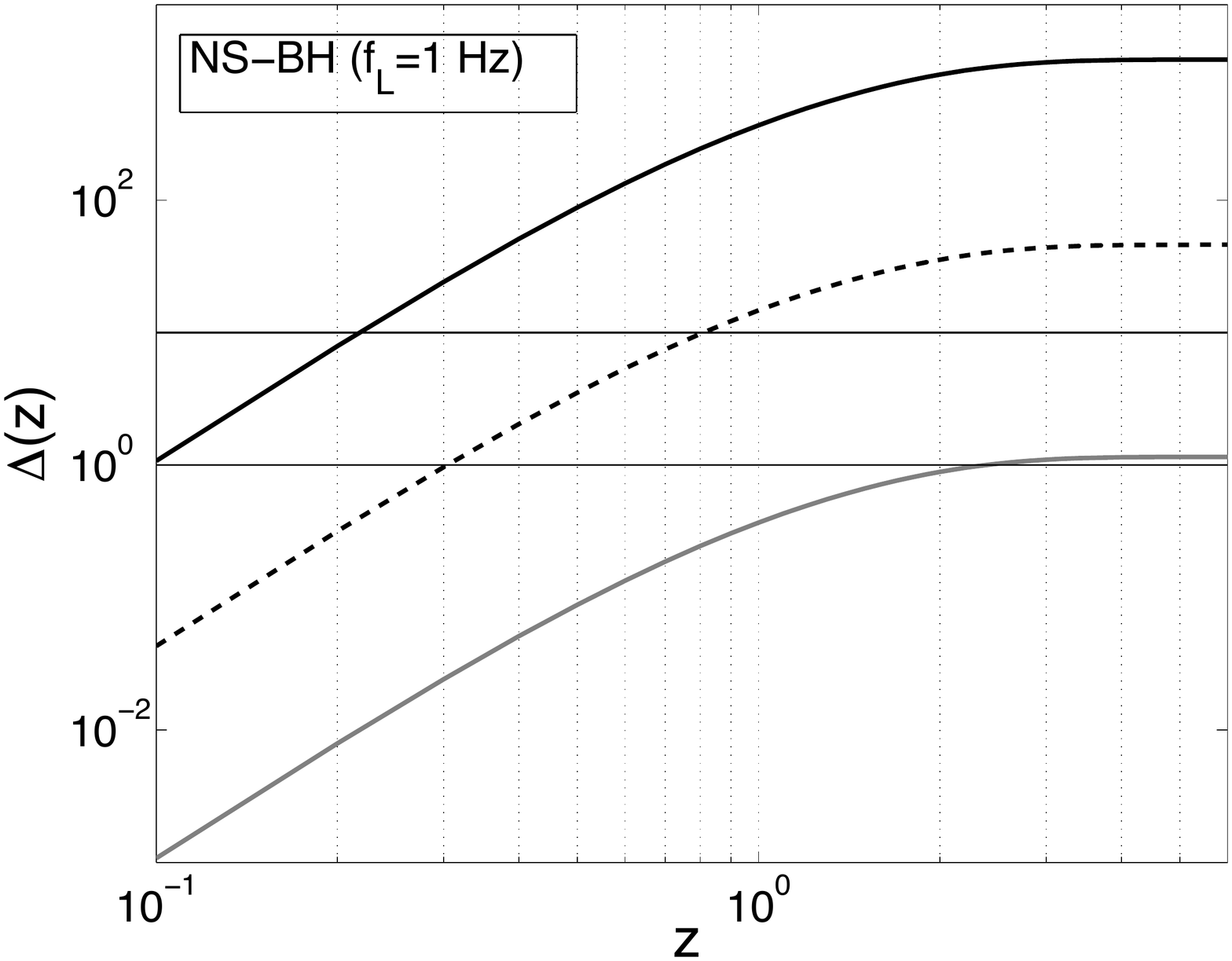}
\caption{Same as Fig.~\ref{fig-DC10} for a lower frequency bound of 1 Hz. 
\label{fig-DC1}}
\end{figure}  

Figure \ref{fig-DC10} shows $\Delta(z)$ for different estimates of the
local coalescence rate of NS-NSs and NS-BHs and for an instrumental
seismic wall of 10 Hz; Figs.\ {\ref{fig-DC5}} and {\ref{fig-DC1}} show
the same thing for seismic walls at 5 and 1 Hz, respectively.  The
threshold redshifts $z_*$ and $z_{**}$ for NS-NS are shown in
Table~\ref{table-zNSNS}; Table \ref{table-zNSBH} shows these
thresholds for NS-BH.  For ease of comparison, we repeat in these
tables the detection horizons for the various instruments discussed
above.

\begin{table}
\caption{\label{table-zNSNS} Top entries: threshold between resolved
and unresolved NS-NS binaries for different estimates of the source
rate $\dot\rho_c^{\rm o}$ and detector lower frequency bound $f_L$. No
value means that the duty cycle is always $<1$ or $<10$.  Bottom
entries: lower frequency sensitivity bound and detection horizon for
current and planned gravitational-wave detectors. For LIGO, the horizon is given for the SRD  4 km detector and the value in parenthesis corresponds to the combination of the three detectors.}
\begin{ruledtabular}
\begin{tabular}{lccc}

$f_L$ & $\dot\rho_c^{\rm o}$ &  $z_*$ & $z_{**}$ \\
10 & 0.01 & - & -\\
    & 0.4 & 0.8-0.9 & -  \\
    & 1 & 0.5-0.6 &  $>2$\\
    & 10 & 0.2 & 0.5-0.6\\
\hline
5 & 0.01 & - & - \\
    & 0.4 & 0.4 & 1-1.2\\
    & 1 & 0.25 & 0.6-0.7\\
    & 10 & 0.1 & 0.25 \\
\hline
 1 & 0.01 & 0.3 & 0.8\\
    & 0.4 & 0.08 & 0.2\\
    & 1 & 0.06 & 0.13\\
    & 10 & 0.03 & 0.06\\   
\end{tabular}
\begin{tabular}{lcc}
LIGO: & $f_L = 40$ Hz & $z_{\rm DH} = 0.0035 (0.005)$ \\
Virgo: & $f_L = 10$ Hz & $z_{\rm DH} = 0.0025$ \\
Advanced LIGO: & $f_L = 10$ Hz & $z_{\rm DH} = 0.045 (0.08)$ \\
Advanced Virgo: & $f_L = 5$ Hz & $z_{\rm DH} = 0.035$ \\
Einstein Telescope: & $f_L = \{5,3,1\}$ Hz & $z_{\rm DH} \simeq 1$ \\
\end{tabular}
\end{ruledtabular}
\end{table}

\begin{table}
\caption{\label{table-zNSBH} Same as Table.~\ref{table-zNSNS}, but
for NS-BH binaries.}
\begin{ruledtabular}
\begin{tabular}{lccc}
$f_L$ & $\dot\rho_c^{\rm o}$, &  $z_*$ & $z_{**} $\\
10 & 0.001 & - & - \\
     & 0.04 & - & -\\
     & 1 & 1.1-1.4 & -\\    
 \hline
5 & 0.001 & - & -\\
     & 0.04 & - &  - \\ 
     & 1 & 0.5 & $>1.6$\\  
 \hline    
  1 & 0.001 & $> 2.3$ & - \\
     & 0.04 & 0.3 & 0.8-0.9\\
     & 1 & 0.1 & 0.2\\    
\end{tabular}
\begin{tabular}{lcc}
LIGO: & $f_L = 40$ Hz & $z_{\rm DH} = 0.007 (0.01)$ \\
Virgo: & $f_L = 10$ Hz & $z_{\rm DH} = 0.0055 $ \\
Advanced LIGO: & $f_L = 10$ Hz & $z_{\rm DH} = 0.09 (0.16)$ \\
Advanced Virgo: & $f_L = 5$ Hz & $z_{\rm DH} = $ 0.07 \\
Einstein Telescope: & $f_L = \{5,3,1\}$ Hz & $z_{\rm DH} \simeq 2$ \\
\end{tabular}
\end{ruledtabular}
\end{table}

\section{\label{sec:conc} Discussion and conclusions}

The main concern of the first and second generations of
interferometric gravitational-wave detectors will be improving
sensitivity to ensure unambiguous first detection of gravitational
waves, followed by measurement with sufficient precision to inaugurate
gravitational-wave astronomy as an observational science.  As the
technology improves and these instruments' sensitivity increases, the
possibility emerges that a confusion background may become an
important effect, ultimately limiting the capabilities of these
instruments.

Referring to Tables {\ref{table-zNSNS}} and {\ref{table-zNSBH}}, we
see that there is little danger of a confusion background impacting
these instruments even into the advanced LIGO era.  The confusion
background only has an impact when a detector's ``reach'' (summarized
by the detection horizon $z_{\rm DH}$) becomes large {\it and} its low
frequency ``wall'' (the cutoff frequency $f_L$) becomes low.  A
distant detection horizon can greatly increase the number of sources
that an instrument can reach; a low frequency ``wall'' can greatly
increase the amount of time that a source spends in band.

The combination of these two effects is summarized by the duty cycle,
$\Delta(z)$, which is the ratio of the typical duration of a measured
event to the typical interval between events.  We find that, for
advanced detectors with excellent low frequency sensitivity (such as
the planned advanced Virgo design), the redshift $z_*$ at which the
duty cycle for NS-NS is unity, $\Delta(z_*) = 1$, may occur close to
the detection horizon for the reference coalescence rate.  The
redshift $z_*$ defines a ``popcorn'' background in the language of
Sec.\ {\ref{sec:DC}}.  The conservative redshift $z_{**}$ for which
$\Delta(z_{**}) = 10$ (defining a Gaussian stochastic confusion
background) may even be close to the detection horizon for optimistic
event rates.

We find that, for NS-NS, both $z_*$ and $z_{**}$ are well within the
horizon of the planned Einstein Telescope, if its low frequency
sensitivity is at $f_L = 1$ Hz (ET1).  If $f_L = 5$ Hz (ET5), we
expect the popcorn background to occur before the detection horizon,
and more likely around $z_* \sim 0.25-0.4$, unless our most
pessimistic coalescence rates are accurate ($\dot\rho_c^{\rm o}
<0.015$ Myr$^{-1}$ Mpc$^{-3}$).  The transition to a Gaussian
stochastic most likely occurs at $z_{**} \sim 0.6 - 1.2$, but can fall
beyond the detection horizon if $\dot\rho_c^{\rm o} <0.15$ Myr$^{-1}$
Mpc$^{-3}$.

Our conclusions for NS-BH binaries are similar to those for NS-NS for
ET1, $z_*$ and $z_{**}$ are both more likely to occur well below the
horizon. For ET5 however there is not likely to be enough sources to
create a Gaussian stochastic background (even a popcorn background),
except for the most optimistic coalescence rates ($\dot\rho_c^{\rm o}
>0.6$ Myr$^{-1}$ Mpc$^{-3}$). As a consequence, NS-BHs may always be
resolved provided that we can separate them with adequate data
analysis strategies in the popcorn regime.

This result motivates very careful analysis of how data from an
Einstein-Telescope-type instrument would be analyzed.  Experience from
the Mock LISA Data Challenges {\cite{babak08}} and ideas developed for
the Big Bang Observatory {\cite{cut06}} prove that disentangling
multiple signals in a gravitational-wave detector's datastream is
certainly possible.  However, beyond the proof of the concept, it is
not clear how many of the lessons from these examples carry over to
the case of ground-based detectors.  In particular, even for an
Einstein-Telescope-type detector, the bulk of the potentially confused
signals would be of low signal-to-noise ratio since most events will
come from the volume of the universe near the detection horizon.  It
would be a valuable exercise to examine how well techniques that
simultaneously fit multiple signals do in the ET regime.

\begin{acknowledgments}

We thank B.\ Sathyaprakash, C.\ van den Broeck and D.\ McKechan for
providing informations about the Einstein Telescope, D.\ Shoemaker,
A.\ Vicere and G.\ Losurdo for helpful discussions about advanced LIGO
and Virgo and I.\ Mandel for sharing a useful document on the
coalescence rates. SAH is supported by NSF Grant No.\ PHY-0449884 and
by NASA Grant NNX08AL42G, and gratefully acknowledges support from the
MIT Class of 1956 Career Development Fund.

\end{acknowledgments}

\end{document}